\documentclass[11pt]{article}%
\usepackage[top=1.2in,bottom=1.2in,left=1.2in,right=1.2in]{geometry}
\usepackage{amsmath}
\usepackage{amsfonts}
\usepackage{amssymb}
\usepackage{graphicx}%
\providecommand{\U}[1]{\protect\rule{.1in}{.1in}}

\begin{document}

\title{Electromagnetism and time-asymmetry\thanks{Thanks to Lee Smolin for helpful
discussions, and to Gordon Belot for helpful comments on an earlier draft.}}
\author{Steven Weinstein\thanks{Email: sweinstein@perimeterinstitute.ca}\\Perimeter Institute for Theoretical Physics\\31 Caroline St N, Waterloo, ON\ \ N2L 2Y5, Canada}
\date{}
\maketitle

\begin{abstract}
It is a commonplace to note that in a world governed by special or general
relativity, an observer has access only to data within her past lightcone (if
that). The significance of this for prediction, and thus for confirmation, does
not however seem to have been appreciated. In this paper I show that what we
regard as our most well-confirmed relativistic theory, Maxwell's theory of
electromagnetism, is not at all well-confirmed in the absence of an additional
assumption, the assumption that all fields have sources in their past. I
conclude that we have reason to believe that there is a lawlike time-asymmetry
in the world.

\end{abstract}

The standard relativistic theories of electromagnetism and gravitation,
Maxwell theory and general relativity, are deterministic theories. Indeed, it
has been shown that relativistic classical field theories of this sort are among the
best examples of deterministic theories that we have \cite{Ear86}. Determinism
means that the state of the universe at a moment determines the state in the
future, and we generally think of the empirical success of these theories as
being of a piece with their ability to predict the future on the basis of
present data. \ In particular, we regard Maxwell's theory of electromagnetism
as extraordinarly well-confirmed, at least in the macroscopic realm, because
we are able to make successful predictions. Yet the theory itself yields
determinate predictions only in the event that one is handed Cauchy data,
which in the case of Maxwell theory means electric and magnetic field
strengths, and charge positions and velocities, on spacelike hyperplanes. But
we have no access to this data, and our predictions are in fact based on
observation of data originating in our past lightcone. On the basis of only
\emph{this} data, we can predict virtually \emph{nothing} from these theories.
Yet we make successful predictions all the same. How?

This situation borders on absurdity, and I will argue here that it indicates
that our understanding of these theories is flawed. The reason we can often
successfully predict the future is that in fact the bosonic \textquotedblleft
field\textquotedblright\ degrees of freedom are not independent from, and in
fact are entirely a function of, the fermionic \textquotedblleft
matter\textquotedblright\ degrees of freedom on the past lightcone. A version of electromagnetic theory which properly incorporates this constraint is a theory which is time-asymmetric.

Consider Maxwell's equations:%
\begin{align*}
\nabla\cdot E  &  =\rho\\
\nabla\cdot B  &  =0\\
\nabla\times B  &  =j+\frac{\partial E}{\partial t}\\
\nabla\times E  &  =-\frac{\partial B}{\partial t}\text{.}%
\end{align*}
Given the electric and magnetic fields $E(x,t)$ and $B(x,t)$ at some point $x
$, in an otherwise empty spacetime $(\rho=0=j)$, we would like to predict the
value of these fields $E(x,t+dt)=E+\frac{\partial E}{\partial t}$and
$B(x,t+dt)=B+\frac{\partial B}{\partial t}$ at the same point an instant $dt$
into the future. To do so, we need $\nabla\times B$ and $\nabla\times E$,
which means that we need to know the rate of change of the fields as we move
in a spacelike direction (since $\nabla\times E$ is constructed from the
spatial derivatives of $E$, and similarly for $B)$. But since information
cannot travel faster than light --- it cannot travel in a spacelike direction
--- the most that we can know is the value of the fields at various points in
or on our past lightcone. This information is utterly useless for our task.

How, then, do we ever make predictions? The answer is that we assume that all
fields have \emph{sources}. Static fields are assumed to be Coulomb fields,
sourced by charges, and radiation is assumed to arise from the acceleration of
these charges in the past. \emph{Given} this assumption -- i.e., given the assumption that
there is no source-free radiation coming in from past null infinity -- we often have a
reasonable chance of predicting the future, because we can often \emph{see}
the sources. This condition, $F_{in}=0$, is sometimes called the Sommerfeld
radiation condition, and it is natural to assume it if we are dealing with an
isolated system, shielded from its environment. The crucial point is that it
must be, and in fact \emph{is}, assumed even for systems which are \emph{not}
isolated. For example, we invariably predict that a dot of light in the sky
around midnight will still be there an hour later, because we are confident
that the light is emanating from a star, a configuration of matter with
reasonably well-understood long-term behavior. An assumption as to the
existence of radiation coming in from infinity, from outside the past
lightcone, is necessary (though in general still not sufficient) for
prediction.\footnote{The cosmic microwave background presents an interesting
case. It's unclear whether this radiation arises from charged sources, yet it
is a constant feature of the sky, and we are inclined on this basis to predict
its continued occurrence. Effectively, then, we could be said to be assuming
not that $F_{in}=0$, but that $F_{in}=f(t)$ is at most a function of time,
invariant in space for constant values of the cosmic time parameter.}

The upshot is that Maxwell theory and other relativistic theories are
supported by experimental test only to the extent that they are supplemented
by some sort of time-asymmetric assumption tying the fields to the charge-current distribution. 
This is, among other things, a
nice example of the role of what are sometimes called {\em auxiliary} assumptions
in theory testing and confirmation. Predictions in Maxwell theory require not
only the Maxwell equations of motion but the additional $F_{in}=0$ condition,
plus of course whatever information we have about the (contingent) properties
of our sources. While the standard way of viewing the experimental
testing and confirmation of Maxwell theory is to take the theory, add Cauchy
data as initial conditions, deduce the future behavior,\ and compare with
observation, our situatedness \emph{in} the world precludes access to the
relevant data. Augmentation with the Sommerfeld radiation condition encodes
the additional assumption that all radiation is emitted by sources, and thus
allows us to make predictions, given that sufficient information about the
sources in our past lightcone actually reaches us.\footnote{There is, of
course, no guarantee that we can access all the relevant information in our
past lightcone. \ E.g., if someone next to me points a laser pointer out into
space, away from me, the light from this event will never reach me, despite
the fact that it originated from a source in my past lightcone.}

The interest of this simple observation is that it implies that the theory we
are actually testing, the theory we regard as confirmed by our successful predictions 
of electromagnetic phenomena, is a theory which is the \emph{conjunction} of Maxwell
theory and the Sommerfeld condition. This augmented version of Maxwell theory is manifestly
time-asymmetric, since the requirement that $F_{in}=0$ is only meaningful in
the presence of a temporal direction. As a result, the configuration of
charged matter and associated fields at a given time is a function solely of
the past distribution of the charged matter, but it is a function of the
\emph{future} distribution of both the matter and the fields (unless we
stipulate additionally that $F_{out}=0$). Note that this condition is arrived
at not as a proposal for explaining the observed time-asymmetry of radiation;
this is a related but distinct issue, sometimes treated by postulating special
initial conditions. Rather, it is a condition which we have assumed all along,
and which is essential to extracting any predictive power whatsoever from
Maxwell theory. \ 

The approach to electromagnetic theory proposed here involves a reduction in the number
of degrees of freedom reminiscent of the Wheeler-Feynman absorber 
theory \cite{WF45}\cite{WF49}.
In both cases, the matter degrees of freedom constrain completely the field degrees of freedom.  However, while the proposed Maxwell-Sommerfeld theory assumes that $F_{in}=0$, absorber theory imposes in addition the condition $F_{out}=0$ (hence the name).  This allows absorber theory to get rid of the self-fields of the charged particles and thus to avoid problems of divergent self-energy, but at the cost of offering a cosmologically implausible theory.  In contrast, the theory proposed here, Maxwell theory augmented by the Sommerfeld condition, leaves the self-fields intact.

If we accept that what we ordinarily regard as our confirmation of Maxwell
theory is really a confirmation of Maxwell theory \emph{plus} the
time-asymmetric Sommerfeld condition $F_{in}=0$, then we are attributing a
fundamental time-asymmetry to at least one of the laws of nature. The second
law of thermodynamics also posits a time-asymmetry, but it is widely (though
not universally) believed that this asymmetry is not fundamental. \ Whether it
might be understood as fundamental in connection with a fundamental
time-asymmetry in electromagnetism is an interesting open question.

Perhaps even more interesting is the possibility that the quantization of
time-asymmetric electromagnetism will be markedly different from the ordinary
quantum electrodynamics which comes through the quantization of Maxwell
theory. Since the field at every point is a function of the charge
distribution on the past lightcone, the time-asymmetric theory appears to have
far fewer degrees of freedom, and thus the Hilbert space should be much
smaller than the ordinary Hilbert space.  Zero-point fluctuations in the
charge degrees of freedom will completely determine zero-point fluctuations in
the field degrees of freedom, suggesting a possible mechanism for explaining
the near-zero value of the cosmological constant, a mechanism which is not
reliant on supersymmetry. This possibility is currently under investigation.

\end{document}